\documentclass[prd,preprint,superscriptaddress,nofootinbib,longbibliography,aps]{revtex4-1}
\usepackage[utf8]{inputenc}
\usepackage{graphicx}
\usepackage{longtable}
\usepackage{amsmath}
\usepackage{color}
\usepackage{amssymb}
\usepackage{subfigure}
\usepackage{microtype}
\usepackage[linktoc=all]{hyperref}
\usepackage{cleveref}
\usepackage{stackengine}
\usepackage{color}
\usepackage{bm}
\usepackage{braket}
\usepackage{slashed}
\definecolor{myblue}{rgb}{ 0.188, 0.478,0.858}

\begin{document}  
\title{Detecting the Stochastic Gravitational Wave Background from Massive Gravity with Pulsar Timing Arrays}

\author{Qiuyue Liang}

\email{qyliang@sas.upenn.edu}
\affiliation{Center for Particle Cosmology, Department of Physics and Astronomy, University of Pennsylvania, Philadelphia, Pennsylvania 19104, USA} 
\author{Mark Trodden} \email{trodden@physics.upenn.edu}
\affiliation{Center for Particle Cosmology, Department of Physics and Astronomy, University of Pennsylvania, Philadelphia, Pennsylvania 19104, USA} 

\date{\today}

\begin{abstract}
 We explore the potential of Pulsar Timing Arrays (PTAs) such as NANOGrav, EPTA, and PPTA to detect the Stochastic Gravitational Wave Background (SGWB) in theories of massive gravity. In General Relativity, the function describing the dependence of the correlation between the arrival times of signals from two pulsars on the angle between them is known as the Hellings-Downs curve. We compute the analogous overlap reduction function for massive gravity, including the additional polarization states and the correction due to the mass of the graviton, and compare the result with the Hellings-Downs curve. The primary result is a complete analytical form for the analog Hellings-Downs curve, providing a starting point for future numerical studies aimed at a detailed comparison between PTA data and the predictions of massive gravity. We study both the massless limit and the stationary limit as checks on our calculation, and discuss how our formalism also allows us to study the impact of massive spin-2 dark matter candidates on data from PTAs.
\end{abstract} 

\maketitle

\section{Introduction}
One of the most exciting recent developments in astrophysics and cosmology has been the direct detection of gravitational waves, and the rapid use of this technique to extend our cosmological observations beyond those made using electromagnetic radiation or neutrinos. Much attention has rightly been paid to the information we can glean from observing the gravitational waves generated by individual compact sources, such as black hole-black hole and black hole-neutron star mergers. These observations are providing crucial insights into how these astrophysical bodies behave, and strict constraints on allowed deviations of the theory of gravity from General Relativity (GR).

Beyond observing individual systems, the totality of gravitational waves produced within our Hubble volume contributes to an overall stochastic background of gravitational waves, with the potential to provide complementary information about astrophysical and cosmological objects, and about the underlying theory that governs the generation and propagation of gravitational waves. This stochastic background is the scientific target of a number of current and upcoming projects, such as the North American NanoHertz Observatory for Gravitational Waves (NANOGrav), the European Pulsar Timing Array (EPTA), and the Parkes Pulsar Timing Array (PPTA). These collaborations use correlations among the precision timings of the arrivals of signals from tens of millisecond pulsars as a way to discover minute perturbations in spacetime as gravitational waves permeate the universe. The signal-to-noise ratio of these observations increases over time, and thus they provide a particularly interesting probe. 

With sensitivities in the tens to hundreds of microHertz, a primary target of these pulsar timing arrays (PTAs) is supermassive black hole binary systems. However, they can also provide novel tests of proposed new physics. Some proposed new sources arise in the matter sector of our theories, including the background of gravitational waves produced by decaying cosmic string loops, and that generated during cosmological phase transitions. On the other hand, new physics can also arise in the gravitational sector of our theories, affecting how gravitational waves are generated, how they propagate through spacetime, and opening up the possibility of the production of entirely new polarization states. Indeed, work has already been carried out exploring the impact of different polarization modes in PTA gravitational wave searches~\cite{Dalang:2021qhu, Qin:2018yhy}. 

In this paper, we consider the potential of PTAs to constrain or potentially discover evidence for stochastic gravitational wave background signals in theories of massive gravity. In particular, we will be interested in comparing the predictions of ghost-free massive gravity, with 5 polarization modes, to that of GR. 

In pure GR, PTA observations can be used to probe the stochastic gravitational wave background (SGWB) in the following way. When a gravitational wave passes, it perturbs the spacetime around the pulsar, and thus changes the frequency of the signal in a way that can be described as a (non-cosmological) redshift. Using pulsar observations, one can then measure the correlation function of the integral of this redshift. Assuming an isotropic SGWB, this correlator can be decomposed into two parts: the energy density, and the overlap reduction function. The energy density part encodes the power spectrum, describing the amplitude as a function of frequency. The overlap reduction function describes the spatial dependence of the correlator as a function of the angle between two arbitrary pulsars. In GR, this shape function was calculated in a particular approximation by Hellings and Downs \cite{Hellings:1983fr}. 

In massive gravity both the energy density and the overlap reduction function receive corrections, partly due to the non-zero mass, but also due to the presence of additional polarization modes. Since the amplitude and frequency-dependence of the energy density largely depend on the source of the gravitational wave, and we are interested in the effects on the SGWB, in this paper we focus on the correction to the overlap reduction function. 

In this paper we first derive an analytical result, which can be directly compared with the Hellings-Downs curve in GR, for all polarization modes in massive gravity. In the case of tensor modes, our general result agrees with that in \cite{Lee:2010cg} up to a normalization factor. We also analyze the corrections from vector and scalar modes in this theory. These results involve a particular approximation, the validity of which we numerically verify for each mode. We then compute this {\it analog Hellings-Downs curve} in two opposite regimes --- the massless limit, and the stationary limit. One of the most model-independent and rigorous constraints on the mass of graviton is that it must be less than $10^{-23}$ eV~\cite{deRham:2016nuf,Shao:2020exw} \footnote{Note, however, that in other situations much tighter constraints may hold~\cite{Gupta:2018hgm}.}, whereas the sensitivity of current PTA measurements is in the $1-100 $nHz frequency range ($\sim 10^{-24} - 10^{-22} $ eV). There are two qualitatively different ways for a graviton with sufficient energy to be detected in this range. The first is if the graviton mass is sufficiently low (or even zero, as in GR) that the graviton is relativistic. The massless limit of our analysis is relevant to this case. The other possibility is that the graviton mass is in the range $10^{-24} - 10^{-23}$eV, in which case the stationary limit of our analysis is then appropriate.  Beyond massive gravity, it is possible that other spin 2 particles, such as ultra-light dark matter, which might comprise the galactic halos surrounding pulsars, may also contribute to any measured PTA signal (see~\cite{Armaleo:2020yml} for a related idea). This is a second situation in which the stationary limit of our calculations is relevant. 

Compared with previous work~\cite{Lee_2008, Gair:2015hra, Qin:2020hfy,Lee:2013awh}, here we focus particularly on the specific case of ghost-free massive gravity,
we investigate more general settings in which such massive spin-2 excitations might be relevant, such as in dark matter applications, and we carry out a detailed analysis of the validity of the Hellings-Downs approximation. 

The structure of this paper is as follows. In Sec.\ref{sec,II}, we review the polarization tensors in the theory of ghost-free linearized massive gravity. In Sec. \ref{sec,III}, we discuss the corrections arising from massive gravity to the signal measured in PTA experiments. We calculate the change in the frequency of the pulses, and obtain the overlap reduction function for all five polarization modes. In Sec. \ref{sec, IV}, we then numerically verify that an important approximation to the overlap reduction function is valid within the frequency range and for the distances of pulsars relevant to current PTA experiments. We analytically calculate expressions for this analog Hellings-Downs curve, compare the result with that in GR, and discuss how the graviton mass and extra polarization states affect the observed signals.
Throughout the paper, we will follow the $ + - - - $ signature convention. We will write $p^2 = 0$ for the electromagnetic signals received on earth from the pulsars, and $k^2 = m^2$ for the massive gravitational wave momentum.  

\section{Polarization tensors for massive gravitational waves}
\label{sec,II}
The starting point of an analysis of massive gravitational waves is to construct the polarization tensors describing the different polarization states of the graviton. To do this, we employ a technique that allows us to write down a form for such tensors using the simpler and more familiar polarization vectors for spin-1 fields.
It is well known that a massive spin-1 field contains three modes --- two transverse modes and one longitudinal mode. Its polarization vectors can be expressed as~\cite{Gleisberg:2003ue, Han:1998sg},
\begin{equation}
\begin{array}{l}
\epsilon_{\mu}^{\pm}(k)=\frac{1}{\sqrt{2}}(0, \cos \theta \cos \varphi \mp i \sin \varphi, \cos \theta \sin \varphi \pm i \cos \varphi,-\sin \theta) \\ 
\epsilon_{\mu}^{0}(k)=\frac{1}{\sqrt{ k^{2}}}\left(|\bm{k}|, k_{0} \hat{\bm{\Omega}}  \right) = \left(\frac{|\bm k|}{m}, \frac{k_0}{m } \hat{\bm{\Omega}} \right) \ , \quad \hat{\bm{\Omega}} = (\sin \theta\cos\varphi, \sin \theta\sin\varphi,\cos\theta)\ ,
\end{array}
\label{eq,polarization spin1}
\end{equation}
where
\begin{equation}
k_{\mu}=\left(k_{0},|\bm {k}| \sin \theta \cos \varphi,|\bm {k}| \sin \theta \sin \varphi,|\bm {k}| \cos \theta\right) = k_0 \left(1, \frac{|\bm k|}{k_0} \hat{\bm \Omega}\right)\ ,
\label{eq,direction}
\end{equation}
with $\hat{\bm{\Omega}} $ the unit vector of the gravitational wave's spatial direction. 

Now, for massive spin 2 fields, the analogous action is the Fierz-Pauli action
\begin{equation}
    S=\int d^{4} x \left[\frac{1}{2} \partial_{\lambda} h_{\mu \nu} \partial^{\lambda} h^{\mu \nu}-\partial_{\mu} h_{\nu \lambda} \partial^{\nu} h^{\mu \lambda}+\partial_{\mu} h^{\mu \nu} \partial_{\nu} h-\frac{1}{2} \partial_{\lambda} h \partial^{\lambda} h+ \frac{1}{2} m^{2}\left(h_{\mu \nu} h^{\mu \nu}-h^{2}\right)\right] \ ,
\end{equation}
describing a total of five propagating polarization modes. Using the polarization vectors for a massive spin-1 field, given above, we can construct the polarization tensors for a massive spin 2 field in the following way~\cite{Gleisberg:2003ue, Han:1998sg}
\begin{eqnarray}
\label{eq,polarization spin2}
\epsilon_{\mu \nu}^{(i)}&=&\left\{\epsilon_{\mu \nu}^{(+2)},\epsilon_{\mu \nu}^{(+1)},\epsilon_{\mu \nu}^{(0)},\epsilon_{\mu \nu}^{(-1)},\epsilon_{\mu \nu}^{(-2)}\right\}\nonumber\\
&= & \left\{\epsilon_{\mu}^{+} \epsilon_{\nu}^{+}, \frac{1}{\sqrt{2}}\left(\epsilon_{\mu}^{+} \epsilon_{\nu}^{0}+\epsilon_{\mu}^{0} \epsilon_{\nu}^{+}\right), \frac{1}{\sqrt{6}}\left(\epsilon_{\mu}^{+} \epsilon_{\nu}^{-}+\epsilon_{\mu}^{-} \epsilon_{\nu}^{+}-2 \epsilon_{\mu}^{0} \epsilon_{\nu}^{0}\right), \frac{1}{\sqrt{2}}\left(\epsilon_{\mu}^{-} \epsilon_{\nu}^{0}+\epsilon_{\mu}^{0} \epsilon_{\nu}^{-}\right), \epsilon_{\mu}^{-} \epsilon_{\nu}^{-}\right\}
\end{eqnarray}
These polarization tensors are transverse $k^\mu \epsilon_{\mu\nu} =0 $, traceless $\epsilon^\mu_{~ \mu} = 0$ and orthonormal $\epsilon^s_{\mu\nu}\epsilon^{s' \mu\nu} = \delta^{s s'}$. Since we are working at the linearized level, we can write the spin 2 piece as a plane wave (see the discussion in \cite{Isi:2018miq} and references therein),
\begin{equation}
\label{eq,hmunucoord}
    h_{\mu\nu}(x) =\frac{1}{2\pi}  \int  d^4 k \frac{2 \delta(|\bm k|^2 - (k_0^2-m^2))}{|\bm k|}  e^{i k x} h_{\mu\nu}(k) = \int_{-\infty}^\infty d f \int_{\text{sky} } d^2 \hat{\bm \Omega } ~ e^{i 2\pi f \left(t- \frac{|\bm k|}{k_0} \hat{\bm{\Omega}}  \cdot \bm x \right) }  h_{\mu\nu}\left(f, \frac{|\bm k|}{k_0} \hat{\bm \Omega }\right) \ ,
\end{equation}
where $2\pi f = k_0$ is the frequency of the spin 2 particle. Note that the $\delta$-function is introduced to impose the on-shell condition $\square h_{\mu\nu} + m^2 h_{\mu\nu}=0 \Leftrightarrow k_0^2 =|\bm k|^2 + m^2$, therefore in the second equality we have integrated over the magnitude of $|\bm k|$. Note also that, compared with the conventions often used in quantum field theory, we have defined the transformed quantity with an extra factor of $|\bm k|$ in the denominator, and have also distributed the factors of $2\pi$ in the definitions of the Fourier transform and its inverse so as to make the final equality as simple as possible. 

Finally we can express $h_{\mu\nu}(k)$ in terms of the polarization tensors, 
\begin{equation}
\label{eq,hdecomposition}
h_{\mu \nu}(k)=\sum_{i} h^{(i)}(k) ~ \epsilon_{\mu \nu}^{(i) }  \ ,
\end{equation}
with $i \in\{0, \pm 1, \pm 2\} $. We denote the helicity $\pm 2$ polarization modes as tensor modes, which are related to the usual definitions of cross ($\times$) and plus ($+$) modes in GR by a rotation. Helicity $\pm 1$ modes are vector modes and the helicity $ 0$ mode is a scalar mode. 
We reserve a comparison with more standard polarization modes in the literature for Appendix \ref{appendix,1} with a specific choice $\theta = \varphi =0$.

\section{The signal}
\label{sec,III}
Having discussed the metric perturbation and its polarization modes, we now want to know how these quantities affect the observed pulsar signal. Suppose a pulsar emits a signal of frequency $\nu_0$ in flat spacetime. Then, if there is a gravitational wave passing between the pulsar and our telescopes (e.g. through the solar system), the measured pulsar frequency $\nu(t)$ will differ from $\nu_0$, leading to an anomalous {\it residual},
\begin{equation}
R(t) \equiv \int_0^t dt^\prime\, \left(\frac{\nu_0-\nu(t^\prime)}{\nu_0}\right)
\end{equation}
in the pulse arrival time. The mean square residual $\braket{R^2(t)}$, defined as
\begin{equation}
    \braket{R^2(t) } = \frac{1}{T} \int_0 ^T R^2(t) dt \ ,
\end{equation}
where $T$ is the time interval over which the observations are made, is the crucial quantity that is measured by the PTA system \cite{Detweiler:1979wn}.

To calculate this observable, we start from the wave vector of the pulse
\begin{equation}
    p^\mu = \bar p^\mu + \delta p^\mu\  ,g_{\mu\nu} p^\mu p^\nu = 0\ ,
\end{equation}
where $ \bar{p}^\mu = \omega_P (1, -\hat{\bm p})$ denotes the null vector in Minkowski spacetime, with $\omega_P$ the pulse frequency at emission and $\hat{\bm p}$ the unit vector pointing toward the pulsar. The light-path trajectory in Minkowski spacetime can be parametrized as 
\begin{eqnarray}
\bar{x}^\mu(\lambda)=\left(t_E+\omega_P\left(\lambda-\lambda_E\right),-\omega_P\left(\lambda-\lambda_E\right) \hat{\bm p}\right)\ .
\end{eqnarray}
In the following, we have chosen $\lambda_P = 0$ at the pulsar emission and $\lambda_E = L/\omega_P$ when the observer on Earth measures the signal. 

The wave vector satisfies the geodesic equation, $ p^\mu \nabla_\mu p^\nu=0\ $, from which we can obtain the linearized geodesic equation, 
\begin{eqnarray}
     \frac{d \delta p^\mu}{d \lambda}=-\Gamma_{\nu \rho}^\mu \bar{p}^\nu \bar{p}^\rho \ ,  \delta p^0 (\lambda)= \int_0^\lambda d\lambda' \left(h_{0\nu,\rho} \bar p^\nu \bar p^\rho - \frac{1}{2} h_{\nu\rho,0} \bar p^\nu \bar p^\rho  \right) + C^0\ ,
\end{eqnarray}
where $C_0$ is the integration constant, chosen such that the frequency measured by an observer in the pulsar's local rest frame is $\omega_P$. Using the Fourier transformation of $h_{\mu\nu}$ as in Eq.\eqref{eq,hmunucoord}, 
As can be seen from \eqref{eq,hmunucoord}, a plane wave metric perturbation propagating in a specific direction $\hat{\bm \Omega}$ is given by
\begin{equation}
\label{eq,fourierhA}
 h_{\mu\nu} \left(t-\frac{|\bm k|}{k_0}\hat{\bm{\Omega}}  \cdot \bm x \right) = \int_{-\infty}^\infty d f  ~ e^{i 2\pi f \left(t- \frac{|\bm k|}{k_0} \hat{\bm{\Omega}}  \cdot \bm x \right) }  \tilde h_{\mu\nu}\left(f, \frac{|\bm k|}{k_0} \hat{\bm \Omega }\right)  \ ,
\end{equation}
We can express $dh_{\mu\nu}/d\lambda$ as
\begin{equation}
    \frac{d h_{\mu\nu} (t-\frac{|\bm k|}{k_0} \hat{\bm{\Omega}} \cdot \bm x)}{d\lambda} =\frac{\partial h_{\mu\nu}}{\partial x^0 } \frac{d x^0}{d\lambda} + \frac{\partial h_{\mu\nu}}{\partial\hat{\bm{\Omega}}  \cdot \bm x } \frac{d \hat{\bm{\Omega}} \cdot \bm x }{d\lambda} = \frac{\partial h_{\mu\nu}}{\partial x^0} \left(\frac{dx^0 }{d\lambda} - \frac{|\bm k|}{k_0}\hat{\bm{\Omega}} \cdot \frac{d\bm x }{d\lambda} \right) = \dot h_{\mu\nu} \nu \left(1+ \frac{|\bm k|}{k_0} \hat{\bm{\Omega}} \cdot \hat{\bm p} \right) \ ,
\end{equation}
where $d\bm x/d\lambda = -\nu\hat{ \bm p}$ is the spatial momentum of the pulsar signal. 
We have 
\begin{eqnarray}
    \delta p^0(P)=C^0, \delta p^0(E)=-\bar{p}^\nu \tilde{h}_{0 \nu} \Delta+\frac{1}{2} \frac{\bar{p}^\nu \bar{p}^\rho \tilde{h}_{\nu \rho}}{\omega_P(1+A ~\hat{\bm{\Omega}} \cdot \hat{\bm p}  )  } \Delta+C^0 \ ,
\end{eqnarray}
where we use $A = \frac{|\bm k|}{k_0} $ for simplicity. 
Due to the presence of $h_{0\mu}$ in the scalar and vector polarization modes, the observers at the pulsar and on Earth, $y^\mu_{P,E}$, do not have the same four-velocity as in the synchronous gauge, where the spatial components vanish, since they must also satisfy the timelike geodesic equation,
\begin{eqnarray}
    \frac{d^2 y^\mu} {d\tau^2} = - \Gamma^\mu_{\alpha\beta} \frac{d y^\alpha }{d\tau } \frac{d y^\beta }{d\tau }\ . 
\end{eqnarray}
For a Minkowski observer, the four-velocity is given by
\begin{eqnarray}
\bar u^\mu = \frac{d y^\mu}{d\tau} = (1,0,0,0)\ .
\end{eqnarray}
Linearizing the timelike geodesic equation, one obtains
\begin{eqnarray}
\delta u^\mu = \left(\frac{1}{2} h_{00},-h_{0 i}-A \frac{\hat{\bm \Omega}^i}{2  } h_{00}\right) \ ,
\end{eqnarray}
where $h_{00}$ and $h_{0i}$ denote the metric perturbations evaluated at the pulsar and the Earth, respectively. We can therefore calculate the observed frequency at the pulsar,
\begin{eqnarray}
 \omega_E=\omega_P+\frac{1}{2} \omega_P \tilde{h}_{00} \Delta\left(1+A~\hat{\bm{\Omega}} \cdot \hat{\bm p}  \right)-\bar{p}^\nu \tilde{h}_{0 \nu} \Delta+\frac{1}{2} \frac{\bar{p}^\nu \bar{p}^\rho \tilde{h}_{\nu \rho}}{\omega_P\left(1+A ~ \hat{\bm{\Omega}} \cdot \hat{\bm p} \right)} \Delta \ ,
\end{eqnarray}
where we have chosen the integration constant $C^0$ such that the observed frequency at the pulsar satisfies, $\omega_P = -g_{\mu\nu}u^\mu_P p^\nu = \omega_P$, which gives
\begin{eqnarray}
C^0 = \frac{1}{2} \omega_P h_{00}(P) \left( 1+ A~\hat{\bm{\Omega}} \cdot \hat{\bm p} \right) \ .
\end{eqnarray} 

The redshift is defined as the fractional variation in the frequency measured at the pulsar and the Earth, 
\begin{eqnarray} 
    z \equiv \frac{\omega_P-\omega_E}{  \omega_P} &=& -\frac{1}{2} \tilde h_{00} \Delta \left(1+A~\hat{\bm{\Omega}} \cdot \hat{\bm p} \right) + \hat p^\mu \tilde h_{0\mu } \Delta - \frac{1}{2} \frac{\hat p^\nu \hat p^\rho \tilde h_{\nu \rho} }{    \left(1+A~\hat{\bm{\Omega}} \cdot \hat{\bm p} \right)}\Delta  \ .
\end{eqnarray} 

The redshift can also be written as a function of frequency as,
 \begin{equation}
  z (t,\hat{\bm \Omega} ) = \int_{-\infty}^{\infty} d f e^{i 2 \pi f t}\left(e^{-i 2 \pi f L\left(1+\frac{|\bm k|}{k_0}\hat{\bm{\Omega}}   \cdot \hat{\bm p}  \right)}-1\right)  \sum_{i}  h^{(i)}\left(f, \frac{|\bm k|}{k_0}\hat{\bm \Omega}\right) F^{(i)} (\hat{\bm \Omega} )\ ,
 \end{equation}
 where we have defined the so-called {\it receiving function} $F^{(i)}(\hat{\bm \Omega} )$ as,
 \begin{equation}
 \label{eq,receivingfn}
 F^{(i)} (\hat{\bm\Omega} ) \equiv - \frac{\hat p^\mu \hat p^\nu}{2 \left(1+ A \hat{\bm{\Omega}}   \cdot \hat{\bm p}  \right)}\epsilon^{(i)}_{\mu\nu}   +  \hat p^\mu \epsilon^{(i)}_{0\mu} - \frac{1}{2} \epsilon_{00}^{(i)} (1+ A \hat{\bm{\Omega}}   \cdot \hat{\bm p} )\ ,
 \end{equation} 
 which describes how the variation in the frequency depends on the metric perturbation. These expressions allow us to isolate the Fourier transform of the redshift as
 \begin{equation}
 z(f,\hat{\bm \Omega} ) =  \left(e^{-i 2 \pi f L\left(1+\frac{|\bm k|}{k_0} \hat{\bm{\Omega}}   \cdot \hat{\bm p}   \right)}-1\right)  \sum_{i} h^{(i)}\left(f, \frac{|\bm k|}{k_0}\hat{\bm \Omega}\right) F^{(i)} (\hat{\bm \Omega} )
\label{zft}
 \end{equation}
 
Since the effect of the stochastic gravitational wave background on the measured redshift of a given pulsar consists of contributions from gravitational waves arriving from all directions, a relevant quantity to calculate is the total redshift
 \begin{equation}
     \tilde z (f) \equiv \int_{S^2 } d^2 \hat{\bm\Omega} ~ z (f,\hat{\bm \Omega}  ) \ .
 \end{equation}

Finally, the observable that is relevant to PTA data is the two-point correlation function of this quantity, $\braket{\tilde z (f) \tilde z(f')} $. We will assume (see, for example the NANOGrav 12.5-year results~\cite{Arzoumanian:2020vkk} or the most recent results from PPTA~\cite{Goncharov:2021oub}) that the frequency-dependence of this power spectrum is independent of any spatial correlations. Using this so-called {\it common-process model}, we can separate the observable into two distinct important pieces.
The first of these is $\braket{h^{(i)2}}$, which is related to the fractional energy density of gravitational waves at a given frequency, $\Omega_{\mathrm{gw}}(|f|) \equiv (3M_{\mathrm P}^2H_0^2)^{-1} d \rho_{\mathrm {gw}}/d\ln f$ via~\cite{Anholm:2008wy},
\begin{equation}
\left\langle h^{(i)*} (f, \hat{\bm \Omega}) h^{(i^\prime) }\left(f^{\prime}, \hat{\bm\Omega}^{\prime}\right)\right\rangle= \frac{3 H_{0}^{2}}{32 \pi^{3}} \delta^{2}\left(\hat{\bm \Omega}, \hat{\bm \Omega}^{\prime}\right) \delta_{i i^\prime  } \delta\left(f-f^{\prime}\right)  \times|f|^{-3} \Omega_{\mathrm{gw}}(|f|)\ .
 \end{equation}
The second piece describes the shape of the signal, and is referred to as the {\it overlap reduction function},
 \begin{equation}
 \label{eq,gammaffull}
\Gamma(|f|)= \beta \sum_{i} \int_{S^{2}} d^2 \hat{\bm \Omega}\left(e^{i 2 \pi f L_{1}\left(1+\frac{|\bm k|}{k_0}\hat{\bm \Omega} \cdot \hat{\bm p}_{1}\right)}-1\right)  \times\left(e^{-i 2 \pi f L_{2}\left(1+\frac{|\bm k|}{k_0} \hat{\bm \Omega} \cdot \hat{\bm p}_{2}\right)}-1\right) F_{1}^{(i)}(\hat{\bm \Omega}) F_{2}^{(i)}(\hat{\bm \Omega}) \ ,
 \end{equation}
where $\beta$ is a normalization factor introduced to impose $\Gamma(|f|) =1$ for coincident, co-aligned detectors. Together, these yield
 \begin{equation}
     \braket{\tilde z (f) \tilde z(f') } =  \frac{3 H_{0}^{2}}{32 \beta  \pi^{3}} \delta^{2}\left(\hat{\bm \Omega}, \hat{\bm \Omega}^{\prime}\right) \delta_{i  i^{\prime}} \delta\left(f-f^{\prime}\right)  \times|f|^{-3} \Omega_{\mathrm{gw}}(|f|)  \Gamma(|f|) \ .
 \end{equation}
In the remainder of this paper, we will almost entirely focus on the overlap reduction function. 

The above analysis is quite general. When the underlying theory is General Relativity, then the only propagating perturbation of the metric corresponds to the spin-2 graviton. However, in more general theories, more excitations, of various spins, may also contribute. In the particular example of massive gravity, scalar, vector, and tensor contributions decouple from one another, since their kinetic terms can be diagonalized~\cite{Hinterbichler:2011tt}, and so we may define an overlap reduction function for each type of perturbation
\begin{equation}
    \label{eq,overlapreduction}
\Gamma_{I}(|f|) = \beta_{I} \sum_{i} \int_{S^{2}} d^2 \hat{\bm \Omega}\left(e^{i 2 \pi f L_{1}\left(1+\frac{|\bm k|}{k_0}\hat{\bm \Omega} \cdot \hat{\bm p}_{1}\right)}-1\right)  \left(e^{-i 2 \pi f L_{2}\left(1+\frac{|\bm k|}{k_0} \hat{\bm \Omega} \cdot \hat{\bm p}_{2}\right)}-1\right) F_{1}^{(i)}(\hat{\bm \Omega}) F_{2}^{(i)}(\hat{\bm \Omega}) \ ,
\end{equation}
where $i \in \text{Modes of type } I $ and $I = T,V,S$, represents tensor, vector or scalar modes respectively. 
The full two-point function is a sum over all contributions and takes the schematic form
\begin{equation}
\label{eq,zzdecompose}
\braket{\tilde z^2} \propto \left(\frac{\Omega_{ T}}{\beta_T}\Gamma_T  + \frac{\Omega_{ V}}{\beta_V}\Gamma_V  +\frac{\Omega_{ S}}{\beta_S}  \Gamma_S \right)  =\frac{\Omega_T}{\beta_T}  \Gamma_T  \left(1 + \frac{\Gamma_V}{\Gamma_T} \frac{\Omega_V}{\Omega_T}  \frac{\beta_T}{\beta_V} + \frac{\Gamma_S}{\Gamma_T} \frac{\Omega_S}{\Omega_T}  \frac{\beta_T}{\beta_S}  \right) \ .
\end{equation}
If we are unable to discriminate among different polarization modes, then any signal detected by PTA would be interpreted as a tensor contribution. With this in mind, it is convenient to define an ``effective" overlap reduction function as,
\begin{equation}
\label{eq,effectiveHD}
    \tilde\Gamma_T = \Gamma_T + \Gamma_V  \frac{\Omega_V}{\Omega_T}  \frac{\beta_T}{\beta_V}  + \Gamma_S  \frac{\Omega_S}{\Omega_T}  \frac{\beta_T}{\beta_S} \ .
\end{equation}

Note, however, that the different polarization modes have different frequency dependencies, and thus $\Omega_V/\Omega_T$ is in general a function of frequency. However, this frequency dependence is limited over the sensitivity range of current PTA observations, and we will therefore approximate $\Omega_V/\Omega_T$ as a frequency-independent quantity.

  \section{The overlap reduction function in massive gravity}
  \label{sec, IV}

In this section, we explore the effective overlap reduction function Eq.\eqref{eq,effectiveHD} in massive gravity. We will separately discuss the behaviors of tensor, vector, and scalar modes, before combining their effects to obtain the total signal.  

Our primary goal is to obtain an approximate analytic expression. In the case of GR this can be done by dropping the exponential terms in the analogous expression to~\eqref{eq,overlapreduction}, leading to the Hellings-Downs curve. We seek to make the same approximation here, neglecting the exponential factors in Eq.\eqref{eq,overlapreduction}, and referring to the simplified quantity as the analog Hellings-Downs curve. This is appealing, since the exponential factors remove any hope of an analytical expression for the overlap reduction function, and also make numerical evaluations of this quantity significantly more challenging. Our strategy will be to demonstrate numerically that it is safe to neglect the exponential terms for each polarization mode, and then to pursue an analytic expression for the simplified quantity. Once we have separately shown that the analog Hellings-Downs curves for each mode are reasonable approximations, we will then drop the exponential factor in Eq.\eqref{eq,effectiveHD} to yield the {\it effective analog Hellings-Downs curve}
\begin{equation}
    \label{eq,effectiveHD2}
    \tilde\Gamma_{0,T} = \Gamma_{0,T} + \Gamma_{0,V}  \frac{\Omega_V}{\Omega_T}  \frac{\beta_T}{\beta_V}  + \Gamma_{0,S}  \frac{\Omega_S}{\Omega_T}  \frac{\beta_T}{\beta_S}~\ .
\end{equation}

In the massless theory, the primary impact of the exponential factor is to introduce a damped oscillatory behavior.
In the massive theory, the situation is complicated by the appearance of $\frac{|\bm k| }{k_0}$, which allows for a new scale in the theory. Therefore, as we will see, there are two different oscillation modes in the overlap reduction function. One is a slow oscillation with frequency $2\pi \frac{|\bm k|}{k_0}$, and the other is a fast oscillation with frequency $2\pi$. When we integrate over the direction vector $\hat{\bm\Omega}$, it is the slow oscillation that exhibits the damping behavior seen in GR. As we approach the massless limit, where $\frac{|\bm k|}{k_0} \to 1$, the two frequencies coincide, and the overlap reduction function damps rapidly for $fL>10$ (which is the minimum value relevant for PTA observations), rapidly agreeing with the value of the analog Hellings-Downs curve. Therefore we will see that we can safely drop the exponential factors in this case. In the stationary limit, the slow oscillation mode vanishes, since $\frac{|\bm k|}{k_0} = 0$, while the fast oscillation mode is independent of the momentum, and thus independent of the angle between any two pulsars. Thus this does not contribute to the angular dependence of the overlap reduction function, and the analog Hellings-Downs curve is again a reasonable approximation. 

We now demonstrate the above claims numerically, and then compute the analog Hellings-Downs curve for each  polarization mode. After dropping the exponential factor, the integration over sphere in Eq.\eqref{eq,gammaffull} yields an analytical expression, which is our main result of the paper. We then show how each mode contributes to the effective analog Hellings-Downs curve in Eq.\eqref{eq,effectiveHD2}, in order to compare with the original Hellings-Downs curve in GR. We discuss both the massless limit and the stationary limit for tensor modes and vector modes, and mainly focus on the stationary limit for scalar modes. Finally, we combine all the polarization modes together to give an effective analog Hellings-Downs curve, and compare this to the observed data.

\subsection{The tensor modes}

\begin{figure}[h!]
\centering
\includegraphics[scale=0.7]{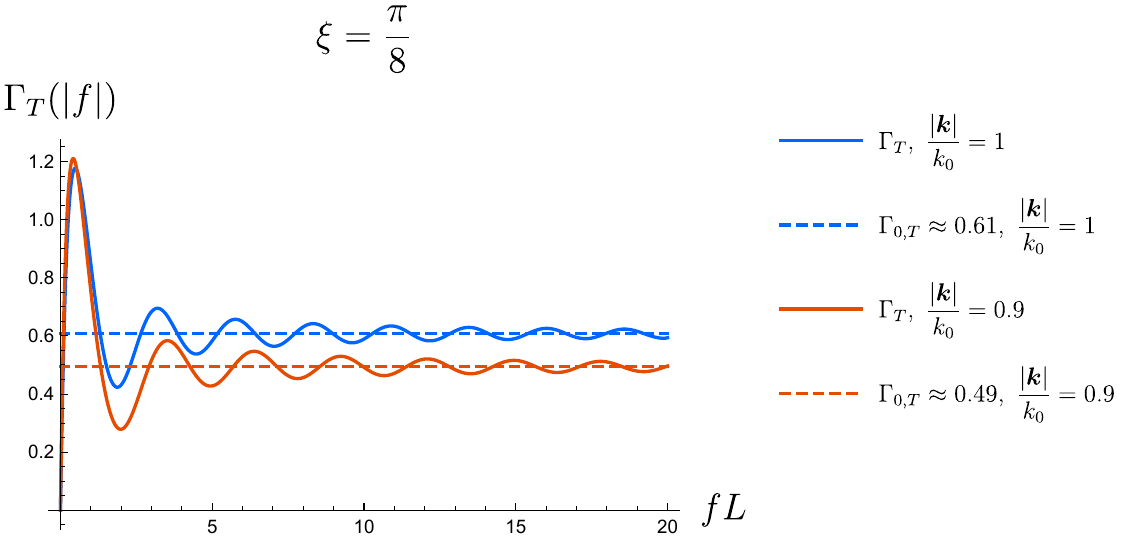} 
 \caption{ Overlap reduction with the exponential factors (solid lines) and without (dashed lines)
for tensor modes. The solid lines are the complete expression (Eq.\eqref{eq,overlapreduction}) as a function of $f L$, and the dashed lines are results of Eq.\eqref{eq,Gamma0T} for $\xi = \pi/8$. The numbers next to the dashed lines in the legend are the values of Eq.\eqref{eq,Gamma0T} for different $\frac{|\bm k|}{k_0}$ and $\xi = \pi/8$. Blue lines are those for the massless case, in which $\frac{|\bm k|}{k_0} =1 $, and red lines represent $\frac{|\bm k|}{k_0} =0.9 $, as an example of tensor modes in massive theories. One can see that for $fL>10$, the deviation between the solid lines and the dashed lines is less than $\sim 5 \%$. 
}
\label{fig:Gammatensor}
\end{figure}

We first evaluate the overlap reduction function for tensor modes, and compare it with the original Hellings-Downs curve. To do this we will ignore the exponential terms in~\eqref{eq,overlapreduction}, and define the resulting simplified quantity as the analog Hellings-Downs curve for tensor modes
 \begin{equation}
 \label{eq,Gamma0T}
     \Gamma_{0,T}=\beta_T \sum_{i =\pm 2} \int_{S^{2}} d \hat{\bm \Omega} F_{1}^{(i)}(\hat{\bm\Omega}) F_{2}^{(i)}(\hat{\bm \Omega}) \ ,
 \end{equation}
 where $F^{(i)}(\hat{\bm \Omega})$ is the receiving function defined in Eq.\eqref{eq,receivingfn}. This is the most direct analogue to the Hellings-Downs curve in massive gravity. 
 In Fig \ref{fig:Gammatensor} we choose a representative value of $\xi$ --- the angle between the two pulsars --- of $\xi = \pi/8$ and compare a numerical evaluation of $\Gamma_T$ to the simplified quantity~\eqref{eq,Gamma0T}, for two values of $\frac{|\bm k|}{k_0}$.  As discussed in~\cite{Anholm:2008wy}, the smallest frequency relevant for PTAs is $f_{\text{min}} \sim 0.1~ \text{yr}^{-1}$, and the closest pulsars used in the observations are at distances around $L_{\text{min}} \sim 100~ \text{ly}$, so that $fL \gtrsim 10$. One can see in Fig~\ref{fig:Gammatensor} that when $fL >10$ the deviation of the full solution (solid curves) from the simplified one (dashed curves) is less than $\sim 5\%$, and so $\Gamma_{0,T}$ is a reasonable approximation to the full overlap reduction function. We have also numerically verified that this conclusion also holds for other values of $\xi$. Thus, for massive tensor modes, and for $|\bm k| /k_0$ close to $1$, dropping the exponential factors in~\eqref{eq,overlapreduction} and instead using~\eqref{eq,Gamma0T} is a valid approximation. 
 
An advantage of adopting the simplified form~\eqref{eq,Gamma0T} of the overlap reduction function is that we can derive an analytic form, for general values of $\xi$. To compute this analog Hellings-Downs curve, we adopt the same coordinate system as that used in~\cite{Anholm:2008wy}, in which we take the spatial part of $\hat p_1$ to be parallel to the $z$ axis, and the spatial part of $\hat p_2$ to lie in the $x- z$ plane, so that we have
 \begin{equation}
     \hat{p}_{1}^\mu=(1,0,0,1),\quad \hat{p}_{2}^\mu=(1,\sin \xi, 0, \cos \xi) \ .
 \end{equation}
The relevant polarization tensors for these tensor modes are then $\epsilon_{\mu\nu}^{(+2)} = \epsilon_\mu^+ \epsilon_\nu^+$ and $\epsilon_{\mu\nu}^{(-2)} = \epsilon_\mu^- \epsilon_\nu^-$ where $\epsilon_\mu^\pm$ are the spin 1 polarization vectors defined in Eq.\eqref{eq,polarization spin1} .  
In fact, these are the exact same polarization tensors as are found in the usual massless case. Substituting these polarization tensors into the receiving function~\eqref{eq,receivingfn}, we see that the second term, proportional to $\epsilon^{(i)}_{0\mu} \hat p^\mu$, vanishes for tensor modes, and so the numerator in the case of massive tensor modes does not change in comparison with the massless case. The expression~\eqref{eq,Gamma0T} then becomes
\begin{equation}
\label{eq,Gamma0T2}
    \Gamma_{0,T} = \frac{\beta_T}{4}\int d^2 \hat{\bm\Omega} \frac{\sin^2 \theta}{2 \left(1+ \frac{|\bm k|}{k_0} \cos\theta \right)} \frac{\cos\xi^2 \sin\theta^2 -2\cos\theta\sin\theta \cos\xi \sin\xi  \cos\varphi + \sin\xi^2 (\cos\theta^2 \cos\varphi^2 - \sin\varphi^2 )  }{1+\frac{|\bm k|}{k_0}  \left( \cos\theta\cos\xi + \cos\varphi \sin\theta\sin\xi\right) } \ .
\end{equation}
Importantly, this differs from the massless case by the presence of the terms proportional to $|\bm k|/k_0$ in the denominator. 
Thus, as expected, in the massless limit, in which $|\bm k|/k_0 \to 1$, $\Gamma_{0,T}$ reverts to the original Hellings-Downs curve \cite{Hellings:1983fr}. 

The details of performing this integration are not particularly illuminating, and so we relegate them to Appendix \ref{appendix,HDcal}. The relevant result is
\begin{eqnarray}
\label{eq,overlapfunctionpm2}
 \Gamma_{0,T} &=& \frac{-\pi}{6A^5 }\frac{\beta_T}{4} \left(4 A  \left( -3 + \left(-6+ 5A^2\right)\cos\xi\right) + 12 \left(1+ \cos\xi + A^2 (1-3\cos\xi) \right)\log \frac{1+A}{1-A}\right. \nonumber\\
 && ~~~  \left. +\frac{3\left(1+2A^2 (1-2\cos\xi) -A^4 (1-2 \cos\xi^2)\right) \log L_1  }{\sqrt{(1-\cos\xi)\left(2- A^2  (1+\cos\xi)\right)}}  \right)
\end{eqnarray}
where we have defined $A = \frac{|\bm k|}{k_0}$ for notational convenience, and where
\begin{equation}
\label{eq,L1}
    L_1 \equiv \frac{\left(1+2 A^2 (1-2\cos\xi) -A^4 (1-2\cos\xi^2 ) - 2 A  (1-A^2 \cos\xi) \sqrt{(1-\cos\xi)\left(2- A^2  (1+\cos\xi)\right) }~ \right)^2}{\left(1- A^2 \right)^4}
\end{equation}

In its full generality, this is a rather complicated expression. However, two limits are of phenomenological interest. One of these is the ultra-relativistic (massless) limit where $A = |\bm k|/k_0 = 1-\epsilon$, with $\epsilon \ll 1$. Expanding to first order in $\epsilon$, we write $\Gamma_{0,T}=\Gamma_{0,T}^{(0)} +\Gamma_{0,T}^{(1)} +\mathcal{O}(\epsilon^2)$, with
\begin{equation}
    \Gamma_{0,T}^{(0)} = \frac{\beta_T}{4}\frac{2\pi }{3} \left(3+ \cos\xi + 6(1-\cos\xi)\log \frac{1-\cos\xi}{2}\right) \ ,
\end{equation}
the usual Hellings-Downs result, as mentioned earlier, and the leading order correction given by
\begin{eqnarray}
   \Gamma_{0,T}^{(1)} = \frac{\beta_T}{4} \frac{2\pi \epsilon}{3} \left(15+11\cos\xi +12(2-\cos\xi) \log \frac{\epsilon}{2} - 3(2-\cos\xi)\log L_1\right)
\end{eqnarray}
While this correction may prove useful in future work, we will not focus on it in this paper, requiring only the leading order result to make our main points.

A second interesting limit is the stationary one, in which $|\bm k|/ k_0 \ll 1 $.  This limit may be relevant for the proposal of massive gravity in some allowed parameter regimes, but there are also other settings in which a massive spin-2 field might contribute to the measurements made by PTAs. One example is the idea of ultra-light spin-2 dark matter, in which the relevant particles would comprise the dominant component of galactic halos. The dark matter mass in such models is such that it may be relevant to the typical frequency range --- $\sim 10^{-9} - 10^{-7}$ Hz $=10^{-24} - 10^{-22}$ eV --- that PTAs are sensitive to. 

\begin{figure}[h!]
\centering
\includegraphics[scale=0.7]{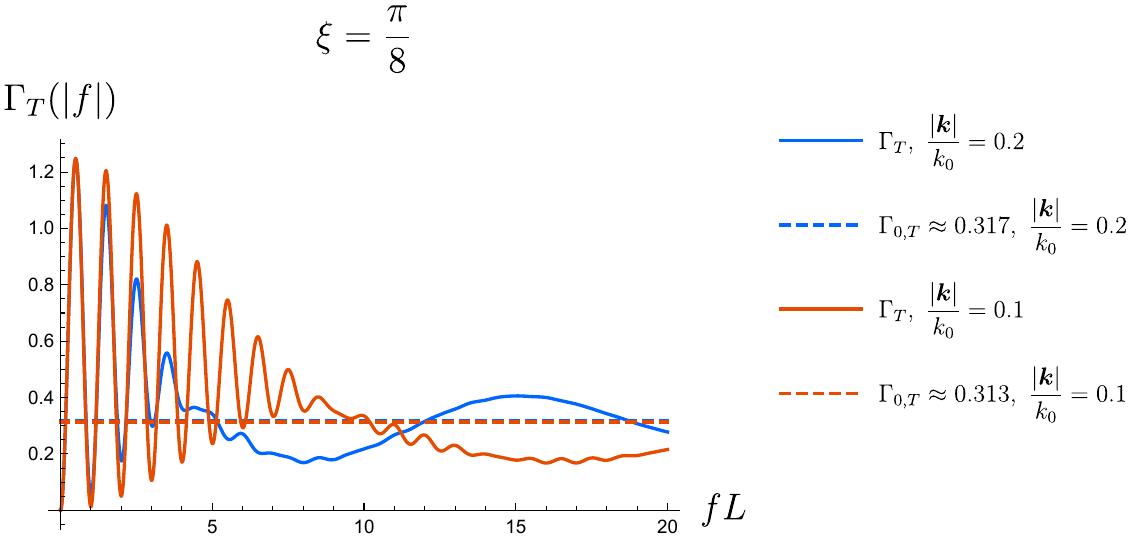} 
\caption{ Overlap reduction with the exponential factors (solid lines) and without (dashed lines) for massive tensor modes for various values of $0\lesssim |\bm k|/ k_0 \ll 1 $. The solid lines are the complete expression (Eq.\eqref{eq,overlapreduction}) as a function of $f L$, and the dashed lines are results of Eq.\eqref{eq,Gamma0T} for $\xi = \pi/8$.  Oscillations are again damped, as in the $|\bm k|/k_0 \sim 1$ case, but here the damping effect is significantly less effective as $|\bm k|/ k_0$ decreases. The fast oscillation can be seen as wiggles upon the slow oscillation, which will dominate when taking smaller $|\bm k|/k_0$. For $fL>10$, the deviation between the solid lines and the dashed lines is less than $\sim 25 \%$. 
} 
\label{fig:Gammatensor2}
\end{figure}

In this limit, $0\lesssim |\bm k|/ k_0 \ll 1 $, it is also possible to show that it is a reasonable approximation to drop the exponential factors in $\Gamma_T$. In Fig \ref{fig:Gammatensor2} we again choose a representative value the angle between the two pulsars of $\xi = \pi/8$ and compare a numerical evaluation of $\Gamma_T$ to the simplified quantity~\eqref{eq,Gamma0T}, for two smaller values of $|\bm k|/ k_0$. One can see the two oscillation modes, and that as $|\bm k|/ k_0 \to 0$ the fast oscillation becomes more prominent, while the period of the slow oscillation increases. Since PTA observations take place over an interval short compared to this period, the slow oscillation can be safely neglected in our approximation. Indeed, when $fL >10$ the deviation of the full solution (solid curves) from the simplified one (dashed curves) is noticeable, but less than a $\sim 25\%$ effect. The fast oscillation, on the other hand, is independent of the angle, and therefore does not complicate the integral. 

We conclude that $\Gamma_{0,T}$ is a reasonable, but by no means perfect, approximation to the full overlap reduction function, as long as $|\bm k|/ k_0$ is not too small. We will find this approximation useful for carrying out a simplified comparison to data and for providing analytic insights into the behavior of the relevant quantities. This is in large part because numerically integrating $\Gamma_T$ for many values of $\xi$ is computationally challenging. However, it is clear that a full numerical evaluation will be needed for precise predictions. We have also numerically verified that this conclusion also holds for other values of $\xi$. 

It is worth noting that when $\bm k/k_0 =0$, there is no damping effect at all, since the exponential factors act as pure oscillation terms $2 - 2 \cos(2\pi fL)$. In this paper, we are particularly focused on the angular dependence, so we focus on the expression~\eqref{eq,Gamma0T}, which we have referred to as the   analog Hellings-Downs curve. The analytical expression in the stationary limit is,
\begin{equation}
   \Gamma_{0,T} =  \frac{\beta_T }{4}\left(\frac{8\pi}{15}\left(-1+ 3 \cos \xi^2 \right)  +
   \frac{8\pi}{105} \frac{|\bm k|^2}{k_0^2} ( -2 - 3\cos\xi + 6 \cos\xi^2 +5\cos\xi^3)\right) + \mathcal{O}\left( \frac{|\bm k|^4}{k_0^4} \right) \ .
\end{equation}  

\begin{figure}[h!]
\centering
\includegraphics[scale=0.8]{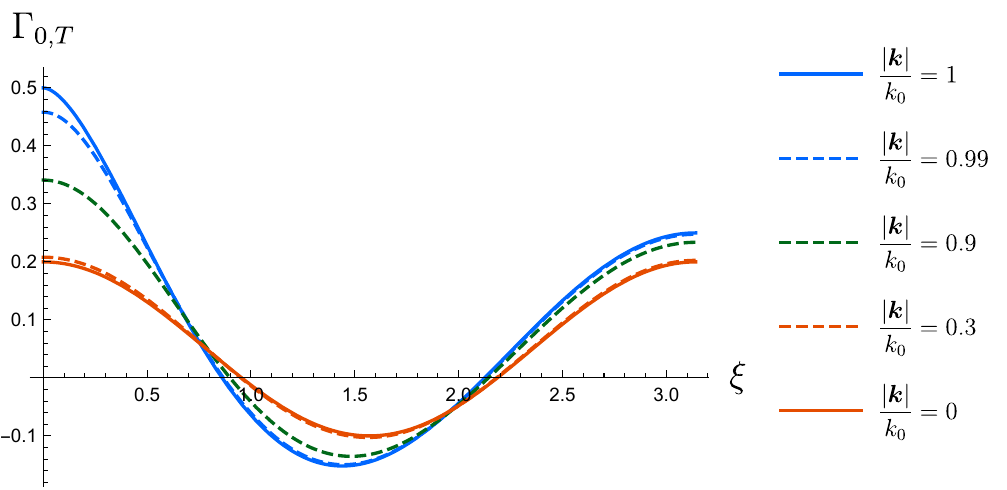} 
\caption{Analog Hellings-Downs curve for tensor modes in massive gravity. The normalization factor $\beta_T$ has been chosen to fix $\Gamma_{0,T} = 0.5$ at $\xi = 0$ in agreement with the massless case. 
The blue curve is the massless original Hellings-Downs curve, while the other lines are the analog Hellings-Downs curves for different values of $\frac{|\bm k|}{k_0}$. The red line is the stationary limit where $\frac{\bm k }{k_0} = 0$.
One can see that the analog Hellings-Downs curve for tensor modes has a suppression at $\xi =0$ and  at $\xi = \pi$. 
} 
\label{fig:tensormassive}
\end{figure}

To obtain more insight into how the graviton mass affects the angular dependence of tensor modes, in Fig.\eqref{fig:tensormassive} we plot the analog Hellings-Downs curve Eq.\eqref{eq,overlapfunctionpm2} for different values of $|\bm k|/k_0$. 

One can see that even a small deviation from the exact massless limit corresponding to GR leads to a visible correction to the curve $\Gamma_{0,T}$. Note that, in Fig.\eqref{fig:tensormassive}, we have chosen the normalization factor $\beta_T = \frac{3}{4\pi}$ for all the different selected values of $|\bm k|/k_0$. If we had chosen a different normalization factor so as to fix $\Gamma_{0,T} = 0.5$ at $\xi = 0$, then we would have recovered the numerical result of~\cite{Lee_2008}.

\subsection{The vector modes}   
  
We now turn to vector modes. As in the tensor case, to simplify our numerical calculations, we first seek to understand whether we can neglect the exponential factors in computing the overlap reduction function. 

We first note that Eq.\eqref{eq,polarization spin2} implies that the polarization modes for vector modes are $\epsilon_{\mu\nu}^{(\pm 1)} = \frac{1}{\sqrt{2}} (\epsilon_\mu^\pm \epsilon_\nu^0 + \epsilon_\nu^\pm \epsilon_\mu^0 )$. 
For simplicity, we denote $\epsilon_\mu^0 = \frac{k_0}{m } \left(\frac{|\bm k|}{k_0}  , \hat{\bm \Omega} \right)$ in the following context. We then use these polarization tensors and the receiving function Eq.\eqref{eq,receivingfn} to obtain the analog Hellings-Downs curve for vector modes as,
\begin{eqnarray}
\label{eq,gamma0V1}
&&   \Gamma_{0,V} = \frac{\beta_V(1-A^2) }{64}  \int d^2\hat{\bm\Omega} \frac{  1 }{ \left(1+ A  \cos\theta \right)\left(1+A   \left( \cos\theta\cos\xi + \cos\varphi \sin\theta\sin\xi\right) \right) }\nonumber\\
  &\times&  \left[  2 \sin^22 \theta + 6 \cos2 \xi\sin^22 \theta + 
 4 \sin^2 2 \theta \sin^2\xi    -8 \cos2\theta \cos\phi  \sin2\theta \sin2\xi   -8 \cos^2\phi    \sin^2 2\theta  \sin^2\xi  \right] \ . \nonumber\\
\end{eqnarray} 
Finally, we evaluate this using the same method as we used for the tensor modes, to yield,
 \begin{eqnarray}
 \label{eq,massivevector}
  &&\Gamma_{0,V}     =  \frac{\beta_V \pi (1-A^2) }{3A^5 }   \Bigg[2A(-3+2(-3+A^2)\cos\xi)+ 6  (-1-(1-A^2)\cos\xi) \log \frac{1-A}{1+A}  \nonumber\\
  && - \frac{3(-1+A^2\cos\xi)}{\sqrt{2-2\cos\xi-A^2 \sin^2\xi}}\log \frac{1-A}{1+A}  + \frac{3(-1+A^2\cos\xi)}{\sqrt{2-2\cos\xi-A^2 \sin^2\xi}} \log L_2  \Bigg] \ ,
 \end{eqnarray}
 where 
 \begin{equation}
  \label{eq,L2}
    L_2\equiv  \frac{1+A-\frac{A^2}{2} -   (1+ A)\cos\xi + \frac{1}{2}A^2\cos2\xi + \sqrt{2} (1-A\cos\xi) \sqrt{2-A^2 -A^2 \cos\xi} \sin\frac{\xi}{2}}{1-A-\frac{A^2}{2} +   (1- A)\cos\xi + \frac{1}{2}A^2\cos2\xi + \sqrt{2} (1+A\cos\xi) \sqrt{2-A^2 -A^2 \cos\xi} \sin\frac{\xi}{2} } \ .   
 \end{equation}
  
\begin{figure}[h!]
\centering
\includegraphics[scale=1]{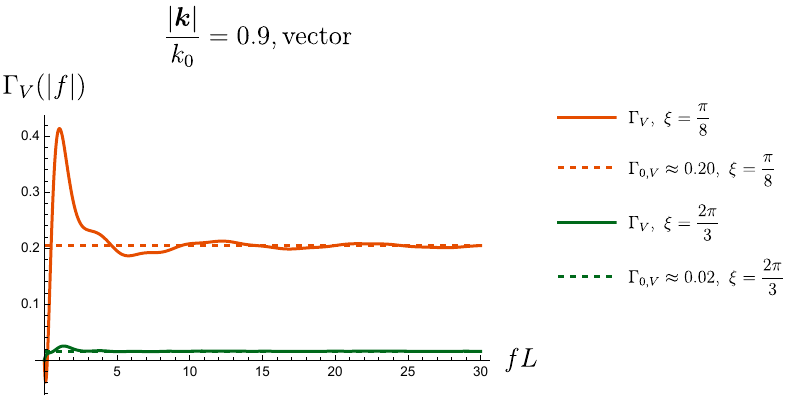} 
\includegraphics[scale=1]{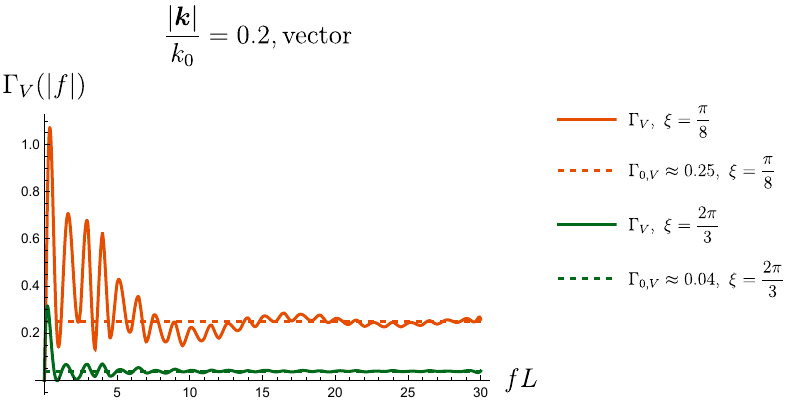} 
\caption{The overlap reduction function and the analog Hellings-Downs curve for vector modes. The solid lines are the full expression, including exponential factors, as a function of $f L$, and the dashed lines are the results of Eq.\eqref{eq,massivevector}. The top panel is for a relatively large value of $|\bm k|/k_0$ and the bottom panel is for a smaller one. In both panels, we plot two examples: $\xi = \pi/8$ (red) and $\xi = 2\pi/3$ (green).} 
\label{fig:Gammavector }
\end{figure}

In Fig.\eqref{fig:Gammavector }, we plot both the full expression Eq.\eqref{eq,overlapreduction} 
and the approximate expression Eq.\eqref{eq,gamma0V1} 
for two different representative values of the angle $\xi$ between the pulsars. We do this separately for $|\bm k| /k_0 = 0.9$ and $|\bm k| /k_0 = 0.2$. In both cases, the overlap reduction function is well-approximated by the analog Hellings-Downs curve in the region $fL>10$ relevant for PTA measurements, and we shall therefore adopt this approximation below. 

\begin{figure}[h!]
\centering
\includegraphics[scale=0.8]{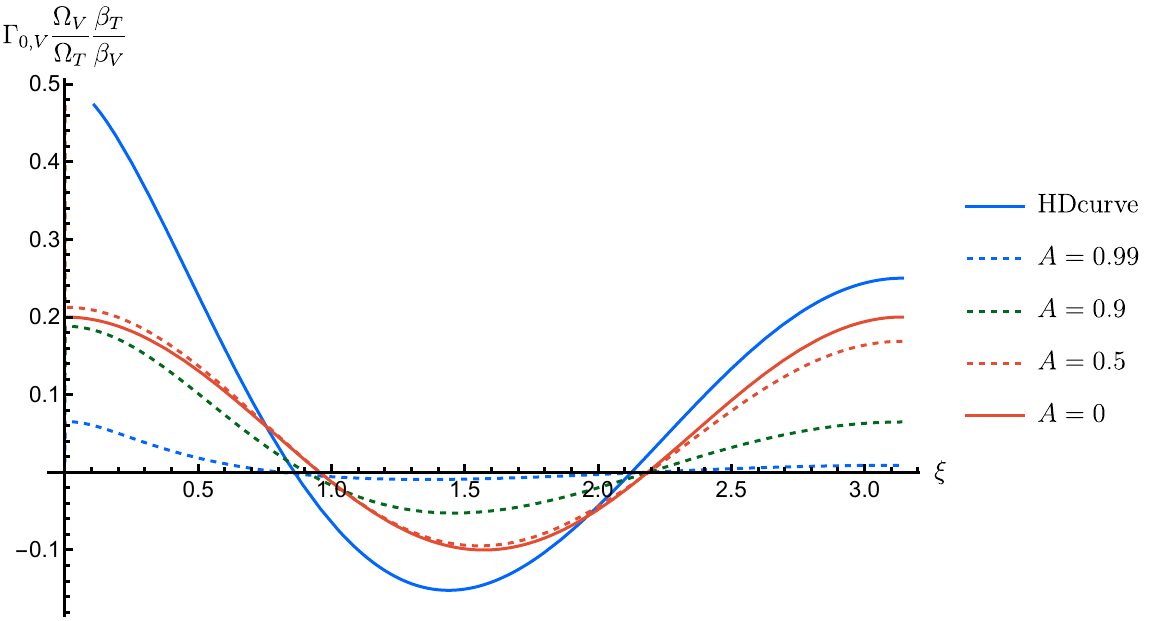} 
\caption{Analog Hellings-Down curve for vector modes in massive gravity. The blue solid curve is the original Hellings-Downs curve in GR, the other curves are $\Gamma_{0,V} \frac{\Omega_V}{\Omega_T}\frac{\beta_T }{\beta_V}$, for different values of $|\bm k|/k_0$, describing the contribution of the vector modes. The plots are made with the parameter $\frac{\Omega_V}{\Omega_T}\frac{\beta_T }{\beta_V} = 1$ to manifestly demonstrate the angular dependence of the vector modes. }
\label{fig:vectormassive}
\end{figure}

To understand what happens in the massless limit, we note that the leading order contribution as $|\bm k|/k_0  \to 1 $  is vanishing, and the next leading order contribution is,
\begin{equation} 
 \Gamma_{0,V} = - \beta_V   \frac{4\pi}{3} (3+4 \cos\xi+ 6 \log \sec\frac{\xi}{2})(1-A)\ .
\end{equation}
The other interesting limit is the stationary one, where $|\bm k|/k_0 \to 0$. In this case the leading order result is
\begin{equation}
    \Gamma_{0,V} =  \beta_V \frac{ \pi}{15}(1+3\cos 2\xi ) \ .
\end{equation}
To obtain a sense of how the vector mode contributes to the analog Hellings Downs curve, we choose $\frac{\Omega_V}{\Omega_T}\frac{\beta_T}{\beta_V} = 1$ in Eq.\eqref{eq,effectiveHD}, and plot the comparison between $\Gamma_{0,V}$ and the original Hellings-Downs curve in GR. 
We show in Fig.\eqref{fig:vectormassive} that for large $|\bm k|/k_0$, the vector modes enhance the curve at $\xi =0$, and suppress it at $\xi = \pi$, whereas for small $|\bm k|/k_0$, the vector modes enhance the analog Hellings-Downs curve at both ends. 

\subsection{The scalar mode } 
\begin{figure}[h!]
\centering
\includegraphics[scale=1]{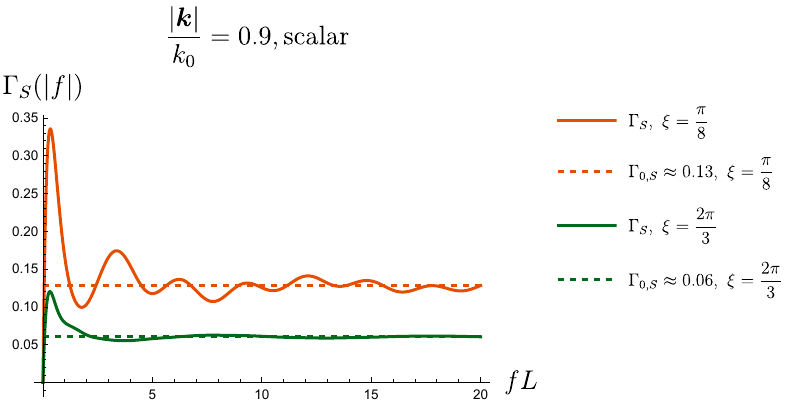} 
\includegraphics[scale=1]{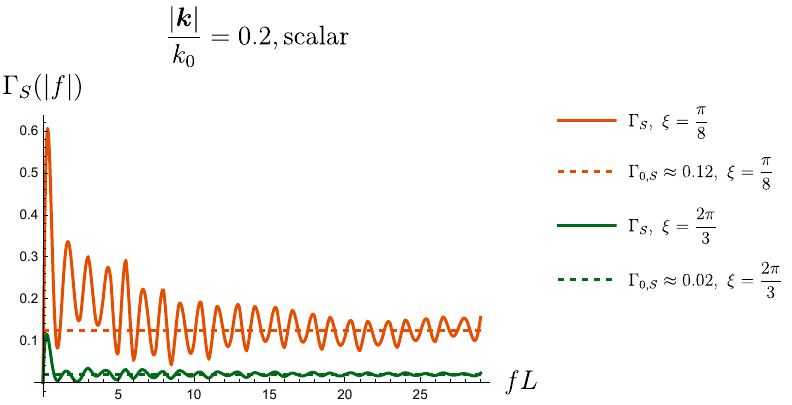} 
\caption{The overlap reduction function and the analog Hellings-Downs curve for scalar mode. The solid lines are the full expression, including the exponential factors, as a function of $f L$, and the dashed lines are results of Eq.\eqref{eq,massivescalar}. One can see that for $fL>10$, it is safe to neglect the exponential factors. } 
\label{fig:Gammascalar }
\end{figure}

The story for scalar modes is very similar to that for vector modes. Again, we define a simplified overlap reduction function, which we refer to as the analog Hellings-Downs curve for the scalar polarization, via 
  \begin{eqnarray}
  \label{eq,gamma0S1}
   \Gamma_{0,S} &= &\frac{\beta_S}{2}\frac{1}{12} \int d^2\hat{\bm\Omega}\frac{ (2(-1+ A^2 )\cos^2 \theta + \sin^2\theta)}{  \left(1+ A \cos\theta \right)\left(1+A \left( \cos\theta\cos\xi + \cos\varphi \sin\theta\sin\xi\right) \right) }\nonumber\\
      &&  \times \left[ (2(-1+A^2)\cos^2\theta\cos^2\xi + \cos^2\xi\sin^2\theta + \sin^2\xi) \right.\nonumber\\
      &+& (-6\cos\theta\cos\xi\sin\theta\sin\xi +A^2\sin2\theta\sin\xi)\cos\phi \nonumber\\
      &+& \left.(-\sin^2\xi+\cos^2\theta\sin^2\xi -2\sin^2\theta\sin^2\xi + 2A^2\sin^2\theta\sin^2\xi)\cos^2\phi \right]  \ .
  \end{eqnarray} 
In Fig.\eqref{fig:Gammascalar } we plot the full scalar overlap reduction function and the analog Hellings-Downs curve for selected values of the angle $\xi$ between the pulsars and the value of $|\bm k|/k_0$. Similar to both tensor and vector modes, we observe fast and slow oscillations and conclude that 
the analog Hellings-Downs curve is a reasonable approximation to the full result for values of $fL$ of interest in PTA observations. We therefore use this approximation from now on.

Evaluating Eq.\eqref{eq,gamma0S1} in the same way yields
   \begin{eqnarray}
   \label{eq,massivescalar}
  \Gamma_{0,S}&=&  \frac{\pi\beta_S }{18A^4}    (-3+2A^2)(-9+6A^2 +(-18+15A^2+ 2A^4)\cos\xi)   \nonumber\\
  &+& \frac{\pi \beta_S   (-1+A^2)\left( (3-3A^2) (-3+A^2) (1+\cos\xi)\sqrt{8-8\cos\xi-2A^ + 2A^2\cos2\xi}\right)}{4A^5\sqrt{2-2\cos\xi -A^2\sin^2\xi}}\log\frac{1-A}{1+A} \nonumber\\
  &+& \frac{\pi \beta_S   3(-1+A^2)^2}{4A^5\sqrt{2-2\cos\xi -A^2\sin^2\xi}}\log L_2    \ ,
 \end{eqnarray}
where $L_2$ is defined in Eq.\eqref{eq,L2}.  

\begin{figure}[h!]
\centering
\includegraphics[scale=0.8]{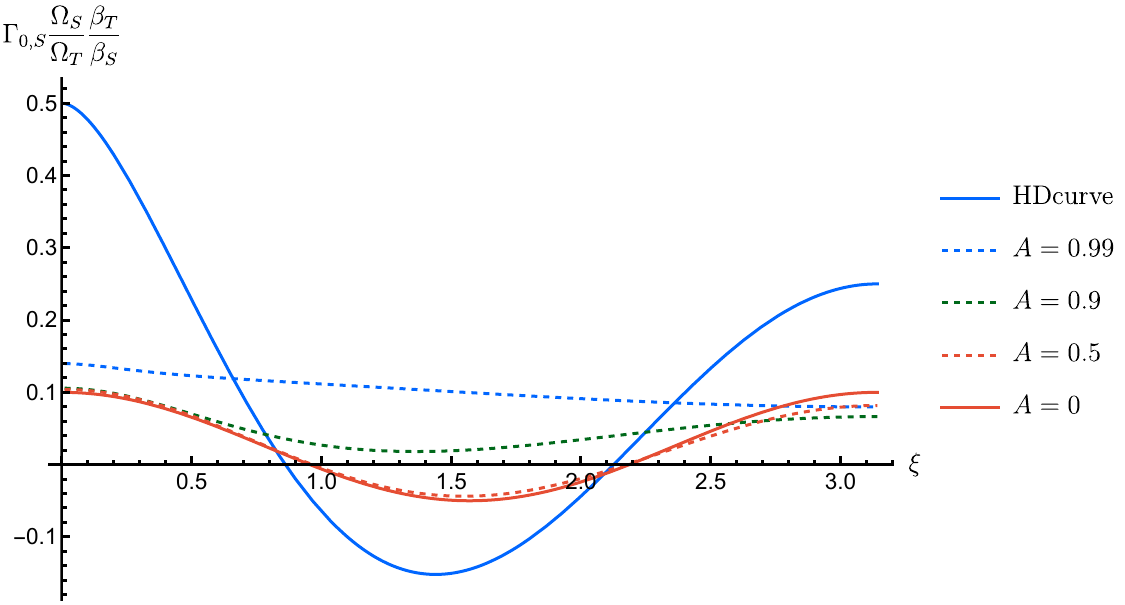} 
\caption{Analog Hellings-Downs curve for scalar mode in massive gravity. The blue solid curve is the original Hellings-Downs curve in GR, the other curves are $\Gamma_{0,S} \frac{\Omega_S}{\Omega_T}\frac{\beta_T }{\beta_S}$, for different values of $|\bm k|/k_0$, describing the contribution of the scalar modes. The plots are made choosing the parameter $\frac{\Omega_S}{\Omega_T}\frac{\beta_T }{\beta_S} = 1$ to show the angular dependence of the scalar mode.  }
\label{fig:scalarmassive}
\end{figure}

We plot the scalar contribution to effective analog Hellings-Downs curve in Fig.\eqref{fig:scalarmassive}, choosing the parameter $\frac{\Omega_S}{\Omega_T } \frac{\beta_T}{\beta_S} = 1$. It is well-known in linearized massive gravity that great care is needed when taking the massless limit in order to treat the scalar mode correctly. This issue is sometimes referred to as the vDVZ discontinuity. Since we do not wish to deviate from the linearized limit, we will not attempt to analyze the massless limit for the scalar mode in this paper. 

The interesting limit we will explore is the stationary limit, where the analog Hellings-Downs curve for the scalar mode is
    \begin{eqnarray}
    \Gamma_{0,S} &= & \frac{\beta_S \pi }{30}  (1+3\cos 2\xi) \ .
   \end{eqnarray}
These results display a similar angular dependence to that of $\Gamma_{0,V}$. Thus, both vector modes and scalar mode contribute in the stationary limit in any theory with massive spin-2 excitations. 
   
\subsection{The combined effective analog Hellings-Downs curve} 

Having separately discussed how each polarization mode contributes to the effective analog Hellings-Downs curve, we now consider the combined effect.

Recall that in Eq.\eqref{eq,zzdecompose} we separated the correlation function into two pieces --- the power spectrum, which encodes its frequency dependence, and the overlap reduction function, describing the spatial angular dependence. Recently, two collaborations --- the NANOGrav 12.5-year result~\cite{Arzoumanian:2020vkk} and the most recent PPTA result~\cite{Goncharov:2021oub} --- have claimed strong evidence for a characteristic power spectrum, but have also claimed that spatial correlations are not well-described by the Hellings-Downs curve. Fitting the power spectrum to a power-law model $f^{-\gamma}$, the PPTA collaboration finds $\gamma \in (1.5, 5.5)$ and NANOGrav collaboration finds $\gamma \in (3.76, 6.78)$. There exists a number of different suggestions for physics that might yield a contribution to the SGWB with a frequency-dependence in this range. Examples are: supermassive black hole binary systems ($\gamma \sim 13/3$) ~\cite{Burke-Spolaor:2018bvk}; primordial gravitational waves ($\gamma\sim 5$)~\cite{Grishchuk:2005qe}; and networks of cosmic strings ($\gamma\sim 16/3$)~\cite{Olmez:2010bi}. It is natural to wonder, therefore, whether modifications of gravity, such as massive gravity, might maintain these successful predictions of the frequency-dependence while modifying the shape of the spatial correlation function, perhaps leading to an improved agreement with current and upcoming data.

In the massive gravity theory that we have studied here, the vector and scalar modes can certainly modify the shape of spatial correlation function. Furthermore, if we ensure that their energy densities are somewhat subdominant to the tensor one ($\Omega_V, \Omega_S \lesssim \Omega_T$), but not so negligible that they do not contribute to the effective analog Hellings-Downs curve, then the frequency-dependence should not be strongly modified. In Fig.\eqref{fig,EffectiveHD}, we have demonstrated the largest this effect could be by choosing the parameters $\Omega_T = \Omega_V = \Omega_S$ in the given frequency band of current PTA observations. One can clearly see that the shape of spatial correlation function can deviate from the Hellings-Downs curve. For some allowed values of the parameters in massive gravity, this deviation can be large enough to be detected as data is accumulated from the current generation of PTA observing programs.  It would be interesting to understand more comprehensively the extent to which upcoming measurements can probe more general modifications of gravity, and massive gravity in particular, using this technique.

 \begin{figure}[h!]
\centering
\includegraphics[scale=0.62]{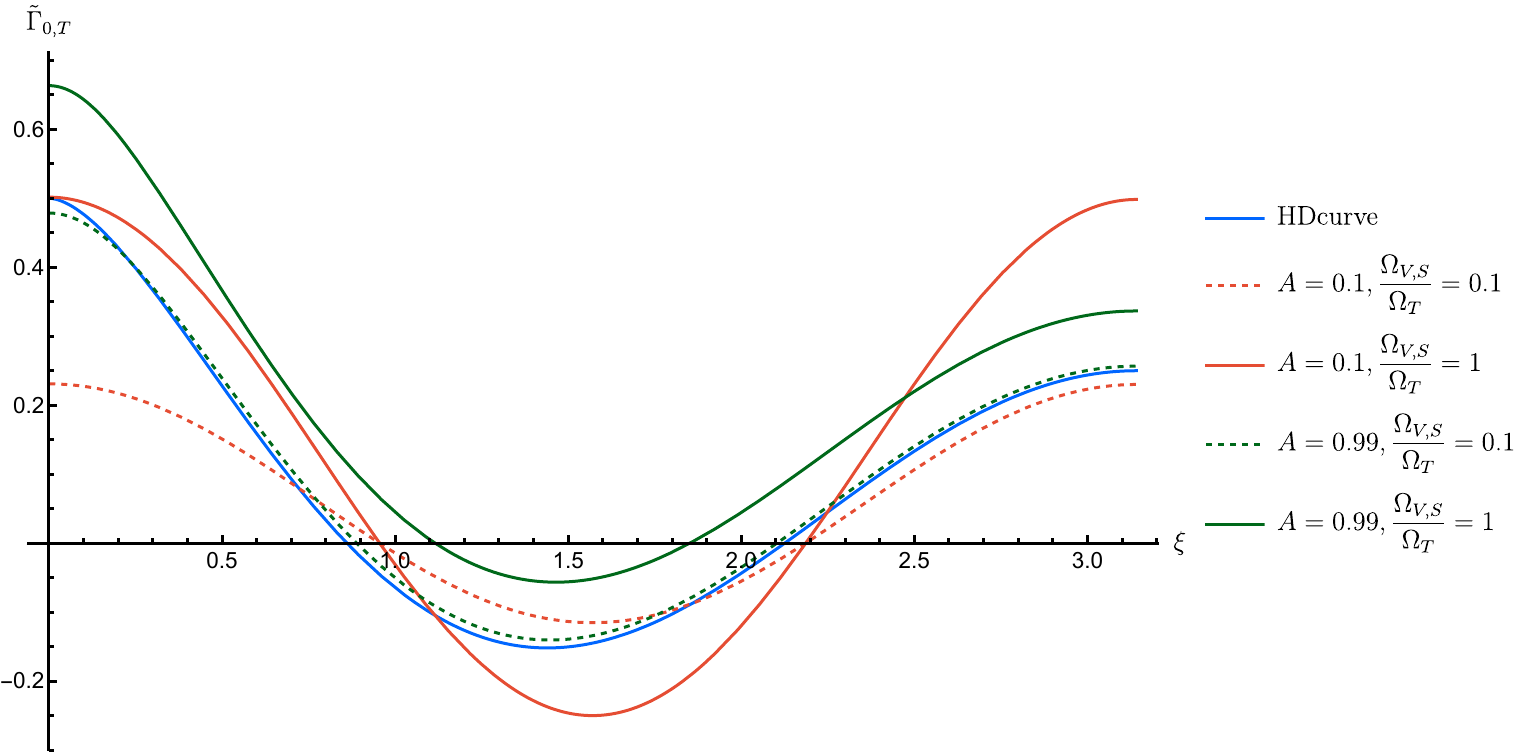} 
\caption{
Comparison between the original Hellings--Downs curve and the effective analog Hellings--Downs curve defined in Eq.~\eqref{eq,effectiveHD2}. The blue line represents the Hellings--Downs curve, the red lines correspond to the non-relativistic limit with $A = 0.1$, and the green lines correspond to the relativistic limit with $A = 0.99$. The solid lines are obtained by assuming that the energy densities stored in the vector and scalar modes are equal to those in the tensor modes, i.e., $\Omega_{V,S}/\Omega_T = 1$, while the dashed lines are chosen to emphasize the tensor modes, with $\Omega_{V,S}/\Omega_T = 0.1$. The normalization factors are chosen to be $\beta_T = \beta_V = \beta_S = 3/4\pi$. } 
\label{fig,EffectiveHD}
\end{figure}

 \section{Conclusions}
In this paper, we have studied the potential of PTA observations to constrain or discover new features of gravity, focusing particularly on the case of massive gravity. PTAs measure the correlation function of arrival times of pulses from pairs of pulsars as a function of the angle between them. Taking the sky-averaged and polarization-averaged product of the response of a pair of Earth-pulsar baselines to a plane wave propagating in a particular direction, the relevant quantity in GR is the famous Hellings-Downs curve, which is an approximation to the overlap reduction function part of the signal's redshift correlation function. Starting with the Fierz-Pauli action that describes linearized massive gravity, we have defined the five  polarization tensors of the relevant metric perturbation around flat spacetime, and have derived how these affect the propagation of signals from the pulsars to our detectors. We have traced how these changes affect the overlap reduction function, and have then defined the analog Hellings-Downs curve for massive gravity.

After numerically justifying the approximations that we make, the
main result of this paper is a full analytical expression for this analog Hellings-Downs curve for the tensor, vector, and scalar modes. We have analyzed the massless limit and the stationary limit of these expressions, and have combined the effects of all the five polarization states together to yield an effective Hellings-Downs curve. These results indicate that it may be possible to distinguish massive gravity from GR as future PTA data is accumulated.

Our hope is that the analytic expressions we have derived will be of use to observers in making detailed comparisons of current and future PTA data with the predictions of theories with massive spin-2 excitations. Should evidence for such massive modes be discovered, a natural question is how such a stochastic gravitational wave background might be generated. In future work we will explore how such signals might arise in theories with massive spin-2 particles, such as massive gravity and spin-2 dark matter models. For example, it would be interesting to study how the interactions of spin-2 dark matter halos can change the effective metric around pulsars, and to understand how massive gravity might give rise to significant gravitational waves through  supermassive black hole binary systems, or phase transitions in the early universe.

 \acknowledgements{
QL thanks Jiewen Chen for helpful discussions, and we thank Kurt Hinterbichler for useful comments on a draft of this paper. This work is supported in part by NASA ATP grant 80NSSC18K0694 and by US Department of Energy (HEP) Award DE-SC0013528.  
}
 
 \appendix
 \section{Appendix} 
 \label{appendix,1}
 In this section, we compare the polarization tensors in Eq.\eqref{eq,polarization spin2} to the more standard definition of polarization tensors in massive gravity theory as has been discussed in \cite{Isi:2018miq}. 

To gain a better understanding of the full five modes in the massive spin-2 theory, we set $\theta = \varphi = 0$, so that the polarization vector of a massive spin-1 particle becomes,
\begin{equation}
    \epsilon_{\mu}^{\pm} = \frac{1}{\sqrt{2}} (0,1,\mp i,0 ),\quad \epsilon_{\mu}^{0} = \frac{1}{m}  (|\bm k|,0 ,0,k_0  ) \ .
\end{equation}
Combining these to obtain the polarization tensor basis yields,
\begin{equation}
    \epsilon_{\mu\nu}^{(+2)} =\frac{1}{2} \left(\begin{array}{cccc}
0 & 0 & 0 & 0 \\
0 & 1 & -i & 0 \\
0 &-i & -1 & 0 \\
0 & 0 & 0 & 0
\end{array}\right),\quad \epsilon_{\mu\nu}^{(-2) } = \frac{1}{2} \left(\begin{array}{cccc}
0 & 0 & 0 & 0 \\
0 & 1 & i & 0 \\
0 &i & -1 & 0 \\
0 & 0 & 0 & 0
\end{array}\right)
\end{equation}
\begin{equation*}
    \epsilon_{\mu\nu}^{(+1)} = \frac{1}{2m }\left(\begin{array}{cccc}
0 &  |\bm k|  & -i  |\bm k|& 0 \\
 |\bm k|  & 0 & 0 &  k_0  \\
-i |\bm k|  &0 & 0 & -i  k_0  \\
0 & k_0  & -i k_0  & 0
\end{array}\right),\quad \epsilon_{\mu\nu}^{(-1) } = \frac{1}{2m}\left(\begin{array}{cccc}
0 &  |\bm k|  & i  |\bm k|& 0 \\
 |\bm k|  & 0 & 0 &  k_0  \\
i |\bm k|  &0 & 0 & i  k_0  \\
0 & k_0  & i k_0  & 0
\end{array}\right)
\end{equation*}

\begin{equation*}
   \epsilon_{\mu\nu}^{(0) } =  \frac{1}{\sqrt{6}}\left( \left(\begin{array}{cccc}
0 & 0 & 0 & 0 \\
0 & 1 & 0 & 0 \\
0 &0 & 1 & 0 \\
0 & 0 & 0 & 0
\end{array} \right)  -\frac{2}{m^2} \left(\begin{array}{cccc}
|\bm k|^2 & 0 & 0 & k_0 |\bm k|  \\
0 & 0 & 0 & 0 \\
0 &0 & 0 & 0 \\
|\bm k| k_0 & 0 & 0 & k_0^2 
\end{array} \right)  \right) 
\end{equation*}

Thus, the metric perturbation around flat spacetime can be expressed in matrix form as,
\begin{equation} 
   h_{\mu\nu}  =   \left(\begin{array}{cccc}
-\frac{2}{\sqrt{6}} \frac{|\bm k|^2 }{m^2} h^{(0)}  & \frac{|\bm k|}{2m} \left( h^{(1)}+  h^{(-1 )}\right) & -i\frac{|\bm k|}{2m} \left( h^{(1)}- h^{(-1 )}\right)  & -\frac{2}{\sqrt{6}} \frac{|\bm k| k_0  }{m^2} h^{(0)}  \\
\frac{|\bm k|}{2m} \left( h^{(1)}+  h^{(-1 )}\right) & \left(h^{(2)}+h^{(-2)}\right) + \frac{1}{\sqrt{6}} h^{(0)} & -i \left( h^{(2)}-h^{(-2)}\right) & \frac{k_0 }{2m} \left( h^{(1)}+  h^{(-1 )}\right) \\
-i\frac{|\bm k|}{2m} \left( h^{(1)}- h^{(-1 )}\right) &-i \left( h^{(2)}-h^{(-2)}\right)  &  -\left(h^{(2)}+h^{(-2)}\right) + \frac{1}{\sqrt{6}} h^{(0)} & -i\frac{k_0 }{2m} \left( h^{(1)}-  h^{(-1 )}\right) \\
-\frac{2}{\sqrt{6}} \frac{|\bm k| k_0  }{m^2} h^{(0)}   & \frac{k_0 }{2m} \left( h^{(1)}+  h^{(-1 )}\right) & -i\frac{k_0 }{2m} \left( h^{(1)}-  h^{(-1 )}\right) & -\frac{2}{\sqrt{6}} \frac{ k_0^2  }{m^2} h^{(0)} 
\end{array} \right)     \ .
\end{equation}
In the massless limit, where only two polarization states remain, one can see that $h_+ = h^{(2)}+h^{(-2)} $ and $h_\times  =-i \left( h^{(2)}-h^{(-2)} \right)$ are the normal definitions of the cross mode and the plus mode, respectively, in GR. 

Following the conventions of~\cite{Isi:2018miq}, where $\beta = \frac{|\bm k|}{k_0}$ and $\alpha \equiv \frac{m}{k_0}$, the metric perturbation is, 
 \begin{equation}
\label{eq,metricpert}
   h_{\mu\nu}  =   \left(\begin{array}{cccc}
-\beta^2 h_l & \beta h_\text{x}  & \beta h_\text{y}   & - \beta h_l  \\
\beta h_\text{x} &h_++ \frac{1}{2} \alpha^2h_l  & h_\times &   h_\text{x}  \\
\beta h_\text{y} &h_\times   &  -h_+ +  \frac{1}{2} \alpha^2h_l   & h_\text{y}  \\
-\beta h_l   &  h_\text{x} & h_\text{y}  & - h_l 
\end{array} \right)    
\end{equation}
  where $\frac{2}{\sqrt{6}} h^{(0)}/\alpha^2 = h_l$, $(h^{(1)} + h^{(-1)} )/2 \alpha = h_\text{x} $, $- i (h^{(1)} - h^{(-1)} )/2 \alpha = h_\text{y} $, $(h^{(2)} + h^{(-2)} ) = h_+  $ and $- i (h^{(1)} - h^{(-1)} ) = h_\times $. It is straightforward to show that, up to an overall factor, the five polarization modes we have defined in \eqref{eq,polarization spin2} are equivalent to those defined in \cite{Isi:2018miq}.
  
 \section{Computation of the Hellings-Downs curve} 
 \label{appendix,HDcal}
In this appendix, for the convenience of the reader, we describe in detail the integration of Eq.\eqref{eq,Gamma0T2}. For simplicity, we denote $\frac{|\bm k|}{k_0} = A$ in the following equations. Writing $x\equiv \cos\theta $, we have
\begin{eqnarray}
\label{eq,gamma0T3}
\Gamma_{0,T} &=&\frac{\beta_T}{4} \int_{-1}^1 dx \frac{1-x^2}{2(1+A x)} \times \nonumber\\
&& \int_0 ^{2\pi} d\varphi  \frac{\cos\xi^2 (1-x^2)-\sin\xi^2  -2x\sqrt{1-x^2} \cos\xi\sin\xi \cos\varphi + \sin\xi^2 (1+x^2) \cos\varphi^2  }{1+A x \cos\xi +A \sqrt{1-x^2 }  \sin\xi \cos\varphi  } \nonumber\\
&=& \frac{\beta_T}{4} \int_{-1}^1 dx \frac{1-x^2}{2(1+A x)} \times \int_0 ^{2\pi} d\varphi \left(C_1\frac{ 1   }{a+b \cos\varphi  }   + C_2  \frac{\cos\varphi}{a+b \cos\varphi }+ C_3  \frac{\cos\varphi^2 }{a+b \cos\varphi } \right) 
\end{eqnarray}
where $C_i$ are independent of $\varphi$, and $\sin\theta = +\sqrt{1-x^2}$, since $\sin\theta $ is always positive for $0<\theta<\pi$. We can simplify the denominator by defining $a \equiv 1+A x \cos\xi$,$b \equiv A \sqrt{1-x^2} \sin\xi$, and noting that $a^2 -b^2 = (1+A x \cos\xi )^2+ A^2(-1+x^2) \sin^2\xi $.  We can then use standard results (see, e.g.~\cite{gradshteyn2007}) to carry out the integration over $\varphi$, yielding:
\begin{eqnarray}
\label{eq,gamma0T4}
&&\Gamma_{0,T} = \frac{\beta_T}{4} \int_{-1}^1 dx \frac{1-x^2}{2(1+A x)} \left( \frac{2\pi \left(\cos\xi^2 (1-x^2) -\sin\xi^2\right) }{\sqrt{ 1-A^2 + A^2 (\cos\xi + x)^2 }} -2x  \cos\xi  \frac{2\pi}{A    } \left(1- \frac{1+A x \cos\xi}{\sqrt{a^2-b^2 }}\right)  \right.\nonumber\\
&&+ \left. \sin\xi^2 (1+x^2) 2\pi\left(\frac{1}{\sqrt{ a^2-b^2 }} -\frac{1+A x \cos\xi}{(A \sqrt{1-x^2} \sin\xi)^2} + \frac{\sqrt{a^2-b^2 }}{(A \sqrt{1-x^2} \sin\xi)^2}\right) \right)
\end{eqnarray}

The integration over $x$ from $-1$ to $1$ is trivial yet tedious. After some simplification, one recovers the result expressed as in Eq.\eqref{eq,overlapfunctionpm2}. The procedures for computing the analogous results for the vector modes, Eq.\eqref{eq,gamma0V1}, and the scalar mode, Eq.\eqref{eq,gamma0S1}, are very similar to the above. 
 
\bibliography{ref} 
\end{document}